\documentclass{article}

\usepackage{PRIMEarxiv}
\usepackage{authblk}
\usepackage{amsmath}
\usepackage{graphicx}
\usepackage{dcolumn}
\usepackage{bm}

\usepackage[utf8]{inputenc}
\usepackage[T1]{fontenc}
\usepackage{mathptmx}
\usepackage{etoolbox}
\usepackage{textcomp}
\usepackage{gensymb}
\usepackage{hyperref}
\usepackage{appendix}
\usepackage{caption}
\usepackage{subcaption}

\title{Designing a validation experiment for radio frequency condensation}

\author[1]{Lanke Fu}
\affil[1]{Princeton University, Princeton NJ}
\author[1]{E. Litvinova Mitra}
\author[2]{R. Nies}
\affil[2]{Princeton Plasma Physics Laboratory, Princeton NJ}
\author[2]{A. H. Reiman}
\author[3]{M. Austin}
\affil[3]{University of Texas, Austin, Texas}
\author[4]{L. Bardoczi}
\affil[4]{University of California, Irvine, CA}
\author[5]{M. Brookman}
\affil[5]{Commonwealth Fusion Systems, Boston, MA}
\author[6]{Xi Chen}
\author[6]{W. Choi}
\affil[6]{General Atomics, La Jolla, CA}
\author[2]{N. J. Fisch}
\author[2]{Q. Hu}
\author[6]{A. Hyatt}
\author[2]{E. Jung}
\author[6]{R. La Haye}
\author[7]{N. C. Logan}
\affil[7]{Columbia University, New York, NY}%
\author[8]{M. Maraschek}
\affil[8]{Max Planck Institute for Plasma Physics, Garching, Germany}
\author[6]{J. J. McClenaghan}
\author[6]{E. Strait}
\author[6]{A. Welander}
\author[2]{J. Yang}
\author[8]{ASDEX Upgrade team}
\date{\today}

\begin{document}

\maketitle

\begin{abstract}
Theoretical studies have suggested that nonlinear effects can lead to “radio frequency condensation”, which coalesces RF power deposition and driven current near the center of a magnetic island. It is predicted that an initially broad current profile can coalesce in islands when they reach sufficient width, providing automatic stabilization. Experimental validation of the theory has thus far been lacking. This paper proposes experiments on DIII-D for testing and refining the theory of the nonlinear effects.
\end{abstract}

\section{\label{sec:intro}Introduction}

The growth of large islands preceeds 95\% of the disruptions in the Joint
European Torus (JET) tokamak with the ITER-like wall. \cite{gerasimov_overview_2020} Although these islands generally appear at the end of a chain of other off-normal events, it appears that it is the island itself that typically triggers the disruption. \cite{devries2014} A statistical analysis of 250 disruptions on JET found a distinct locked mode amplitude at which the plasma disrupted, corresponding to an island width of about 30\% of the minor radius. \cite{deVries2016} A method for automatically stabilizing such islands would clearly be desirable.

Theoretical investigations in the late 1970's showed that radio frequency (RF) waves could be used to drive plasma currents. \cite{fisch_confining_1978,fisch_creating_1980} Subsequent theoretical investigations in the early 1980's showed that it would be feasible, and desirable, to stabilize magnetic islands using those RF driven currents. \cite{reiman83,yoshioka_numerical_1984} Since that time, RF current drive stabilization of magnetic islands has been extensively demonstrated experimentally. \cite{gantenbein_complete_2000,zohm_physics_2001,zohm_phys_plas_2001,leuterer_recent_2003,la_haye_control_2002,prater_discharge_2003,isayama_complete_2000,isayama_achievement_2003}

More recent theoretical studies have found that a nonlinear effect not included in conventional calculations can have a significant impact on island stabilization \cite{reiman_suppression_2018,rodriguez_rf_2019,rodriguez_rf_2020,jin_pulsed_2020,frank_generation_2020,reiman_disruption_2021,jin_two-fluid_2021,nies_calculating_2020,jin_hot_ion_2021,volpe_avoiding_2015,choi_feedforward_2018,petty_complete_2004, bardoczi_direct_2023, prater_stabilization_2007}. “RF condensation” can cause the RF power deposition and driven current to coalesce near the center of a magnetic island, increasing the stabilization efficiency. In particular, an initially broad current profile can coalesce in islands when they reach sufficient width, providing automatic stabilization with no need for external feedback control.\cite{frank_generation_2020,reiman_disruption_2021,jin_hot_ion_2021} On the other hand, failure to properly account for the nonlinear effect in the aiming of the RF ray trajectories can lead to a shadowing effect that causes the energy in the wave to be prematurely depleted, impairing stabilization. The nonlinear effect is predicted to be relevant for ITER. \cite{nies_calculating_2020,reiman_disruption_2021} A recent reactor design makes use of the effect to stabilize tearing modes at modest applied RF powers. \cite{frank_reactor_2022} Experimental validation of the theory has thus far been lacking. In this paper, we will investigate some possible scenarios for testing and refining the theory in a  contemporary tokamak.

In a fusion reactor that takes advantage of the condensation effect, the detailed design of the plasma equilibrium and the locations of the RF launchers can take into account the desired stabilization effect. An experiment in an existing device, however, will be constrained by the available hardware, and it will use a previously studied shot as the reference point for the design of the experiment to minimize the risk of unanticipated difficulties in the running of the experiment. Our calculations in this paper primarily use DIII-D shot 141060. This shot has been previously used to study electron cyclotron current drive (ECCD) stabilization of locked islands \cite{volpe_2015}. In addition, calculations comparing different launcher configurations are are conducted on AUG shot 35350.

The nonlinear effect that is the focus of this paper arises from the sensitivity of the RF power deposition and the RF driven current to a perturbation of the temperature. The temperature perturbation causes an increase in the power deposition in the island, producing a further increase in the island temperature. The nonlinear feedback gives an enhanced temperature perturbation in the island. There is a threshold in the magnitude of the temperature perturbation at which the nonlinear solution of the steady-state thermal diffusion equation bifurcates, giving a discontinuous increase in the temperature perturbation. The electron cyclotron (EC) driven current is itself also sensitive to the temperature perturbation, further enhancing the condensation effect. The nonlinear enhancement of temperature perturbation is the key component of the condensation effect and where the main uncertainties lie. 

\begin{figure*}
    \includegraphics[width=\textwidth]{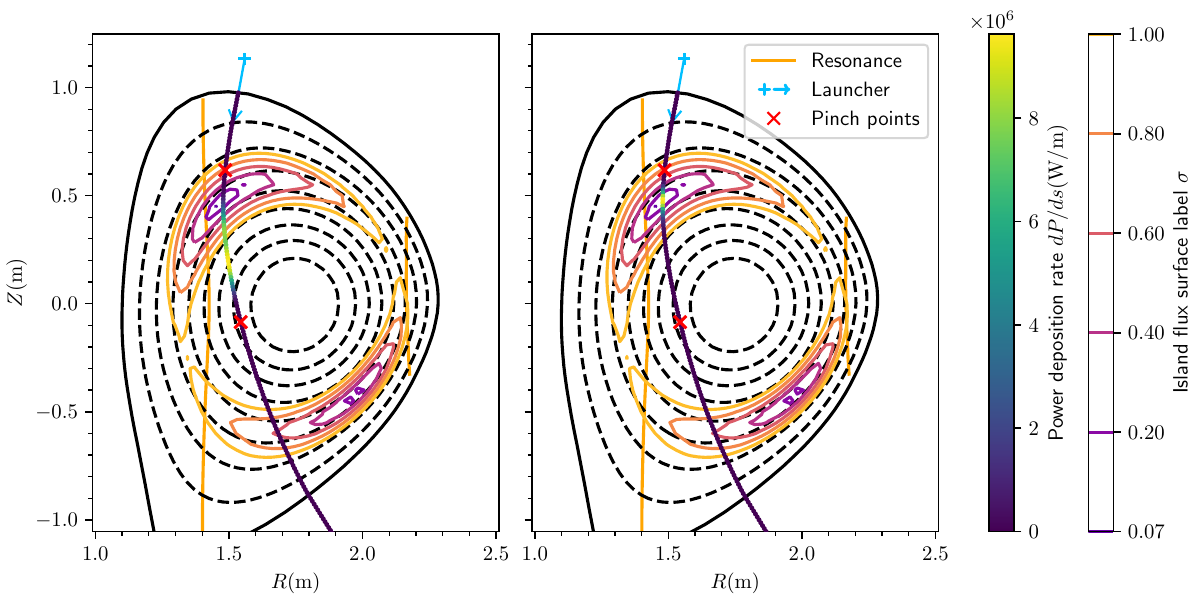}
    \caption{\label{fig:intro_example} Linear (left) and nonlinear (right) solution for the power deposition along a top launch ray trajectory in DIII-D as calculated by the OCCAMI code.}
\end{figure*}

The ECCD efficiency for a given plasma temperature and density has been well validated experimentally. \cite{prater_heating_2004}  This paper will focus on investigating some possible opportunities to test and refine the theory describing the nonlinear  enhancement of the temperature perturbation. To simulate the nonlinear enhancement of the temperature perturbation in magnetic islands, we have used the OCCAMI code. \cite{nies_calculating_2020} The OCCAMI code takes as input a numerically specified axisymmetric equilibrium, and it constructs a 3D field with an island of specified width. \cite{reiman2016} Fig. \ref{fig:intro_example} shows a cross-section through such a 3D equilibrium, including a ray trajectory through the plasma. The GENRAY code \cite{smirnov1994general} is used to calculate the ray trajectories of the launched electron cyclotron waves. The solution for the magnetic field with an island is used to determine the density and temperature profiles along the ray trajectory, which are fed to GENRAY to calculate the power deposition. The code iterates to find the solution to the nonlinear problem defined by the thermal diffusion equation in the island with a source term determined self-consistently by GENRAY. This allows OCCAMI to identify non-linear effects neglected by existing codes.

This paper investigates possible scenarios with EC waves launched above the plasma ("top launch") and with EC waves launched from the low field side of the plasma ("outside launch"). Section \ref{sec:background} provides theoretical background. Section \ref{sec:methods} provides additional details on the OCCAMI simulation code. Section \ref{sec:diagnostics} discusses relevant diagnostic challenges. Section \ref{sec:results} discusses the DIII-D calculations. Section \ref{analysis} examines one of the DIII-D $B_\text{tor}$ scans in greater detail. Section \ref{sec:AUG} discusses the ASDEX Upgrade calculations. Section \ref{sec:discussions} discusses the conditions governing the magnitude of the nonlinear effects as seen in our calculations. Finally, Section \ref{sec:conclusions} discusses the results and some conclusions.

\section{\label{sec:background}Theoretical background}

\subsection{\label{sec:power}Power Deposition}

Electron cyclotron waves in a tokamak generally deposit their energy on the tail of the electron distribution function, where the number of electrons is sensitive to the temperature of the bulk electrons. In a magnetic island, the sensitivity of the power deposition to the temperature perturbation, combined with the local thermal insulation associated with the island, causes a further increase in the temperature. This is the nonlinear feedback effect, discussed in the introduction, that is the key to RF condensation. 

If the EC waves dominantly deposit their power at a phase velocity $V_p$, then the number of resonant electrons, and the power deposition, is roughly given by
\begin{equation}
dP_\text{dep}/ds \propto \text{exp}(-w^2),
\end{equation} 
where $w = V_p/V_T$ and $V_T$ is the thermal velocity. This deposition rate is exponentially sensitive to small temperature perturbations $\widetilde{T}$:
\begin{equation}
dP_\text{dep}/ds \propto \text{exp}(-w_\text{eff}^2)\text{exp}\left(w_\text{eff}^2\frac{\widetilde{T}}{T_0}\right),
\end{equation}
where $w_\text{eff}$ and $T_0$ are the initial values of $w$ and $T$ before the perturbation. This argument has been put on a firmer footing in \cite{nies_calculating_2020}.

The coalescence effect associated with RF condensation is, in some circumstances, strongly constrained by relativistic effects. Those effects constrain the power deposition to lie in the region where:
\begin{gather}
n\Omega/\omega  \geq \sqrt{1-N_{\parallel}^2},\label{eq:relativistic}\\
N_{\parallel} =ck_{\parallel}/\omega.
\end{gather}
Here, $\Omega$ is the electron cyclotron frequency, $n$ specifies the relevant harmonic, $\omega$ is the wave frequency, $c$ is the speed of light $k_\parallel$ is the parallel wave number, and $N_{\parallel}$ is the parallel refractive index.
\cite{prater_heating_2004} The bounding points of the region where this condition is satisfied are called "pinch points". When relativistic effects restrict the power deposition to a narrow region, that can prevent the RF from coalescing. The red crosses in Fig. \ref{fig:intro_example} correspond to the pinch points along the ray trajectory. Power can be deposited only between those pinch points.

\subsection{\label{sec:transport} Heat Transport in the Island}

There is experimental \cite{inagaki_diffusivity_2004,spakman2008,bardoczi_diffusivity_2016} and computational \cite{hornsby2011} evidence that the diffusion coefficient in a magnetic island is much smaller than the ambient thermal diffusion coefficient in the surrounding plasma when the temperature and density gradients in the island are small. This is plausible, as it may be expected that the small gradients would be below the threshold for triggering microinstabilities.

We assume that the transport in the island to be determined by ion temperature gradient (ITG) modes. To reflect this, we use a conventional model for the effect of a ITG threshold on the thermal diffusion coefficient \cite{garbet_stiffness}:
\begin{equation}
\label{stiff_chi_equation}
\kappa_{\perp}=\begin{cases}
      \kappa_0 & (-\frac{R}{T}\frac{d T}{d r} < k_c)\\
      \kappa_0 \left[1+ \frac{\kappa_s}{\kappa_0}\left(-\frac{R}{T}\frac{d T}{d r} - k_c\right)\right] &(-\frac{R}{T}\frac{d T}{d r} \geq k_c)
    \end{cases} %
\end{equation} 
where $\kappa_0$ is the thermal diffusivity below the ITG threshold, $k_c$ is the normalized ITG threshold, $R$ is the major radius and  $\kappa_s/\kappa_0$ is a measure of the stiffness.

For the values of the parameters in Eq. \ref{stiff_chi_equation}, we consider the experimental data presented in \cite{garbet_stiffness}. The values of $k_c$ vary over a relatively narrow range, from 3 to 8, with 5 or 6 being a reasonable average value. The values of $\kappa_s/\kappa_0$ vary over a relatively broad range. As a reasonable average value we take $\kappa_s/\kappa_0$ to be 3.0 in the following. In practice, we will see that the choice of this parameter does not significantly impact our conclusions in this paper. Once the ITG threshold is exceeded, the profiles become sufficiently stiff that it becomes much more difficult to extract useful data concerning RF condensation. Finally, experiments have indicated that this is $\kappa_0$ least an order of magnitude smaller than the thermal diffusivity outside the island, where the diffusivity is $\kappa_0\approx1m^2/s$. We therefore take
$\kappa_0=0.1m^2/s$.

The calculations for this paper have used a single fluid model. Experiments in AUG and W7X have found that, depending on the electron-ion collision frequency, the electron temperature can significantly exceed the ion temperature when the ion temperature is clamped near the ITG threshold. The electrons are affected indirectly by the ITG clamping through their coupling with the ions. This suggests that experiments may produce stronger non-linear effects than predicted by our calculations.

\subsection{\label{sec:top v. side} Launcher location and impacts}
This paper covers possible scenarios with EC waves launched above the plasma ("top launch") and from the low field side of the plasma ("outside launch"). 

Most tokamak experiments with EC waves have been done with outside launch. Outside launch current drive efficiency is highest at intermediate toroidal launch angles, and the capability of outside launchers to launch waves at large toroidal launch angle tends to be limited. (See Section VII and Fig. 13 for a further discussion of these issues.)

Although technically more difficult to implement, top launch ECCD has higher current drive efficiency than outside launch, and therefore more likely to be used in tokamak reactors, to reduce the recirculating power. DIII-D has completed top launch ECCD experiment at large toroidal launch angles. \cite{chen2021doubling,chen_doubling_2022} For the purposes of this study, top launch is of particular interest because the nonlinear effects are the most pronounced at large toroidal launch angles. 

The upper EC launcher on ITER may be regarded as intermediate between top launch and outside launch, with the EC beam launched from above the plasma, but from a major radius larger than that at the magnetic axis ($R=7$m, $z=4.2$ and $4.4 $m).

\section{\label{sec:methods}The OCCAMI simulation code}

Our simulations employ the numerical code OCCAMI (Of Current Condensation Amid Magnetic Islands) \cite{nies_calculating_2020}, which iterates between the GENRAY geometrical optics ray tracing code \cite{smirnov1994general} to evaluate the island power deposition, and a diffusion equation solver for the island temperature profile. The code uses a numerically specified equilibrium file produced by the EFIT code. \cite{lao1985EFIT} The rational surface of interest and island width are specified as input parameters. A 3D field with an island of the specified width is constructed. \cite{reiman2016}  The density and temperature are initially flattened in the island region. The density profile is kept flat. The code iterates to calculate a self-consistent temperature profile in the island with the heat source term calculated by GENRAY.

\begin{figure}
\centering
\includegraphics[width=0.35\textwidth]{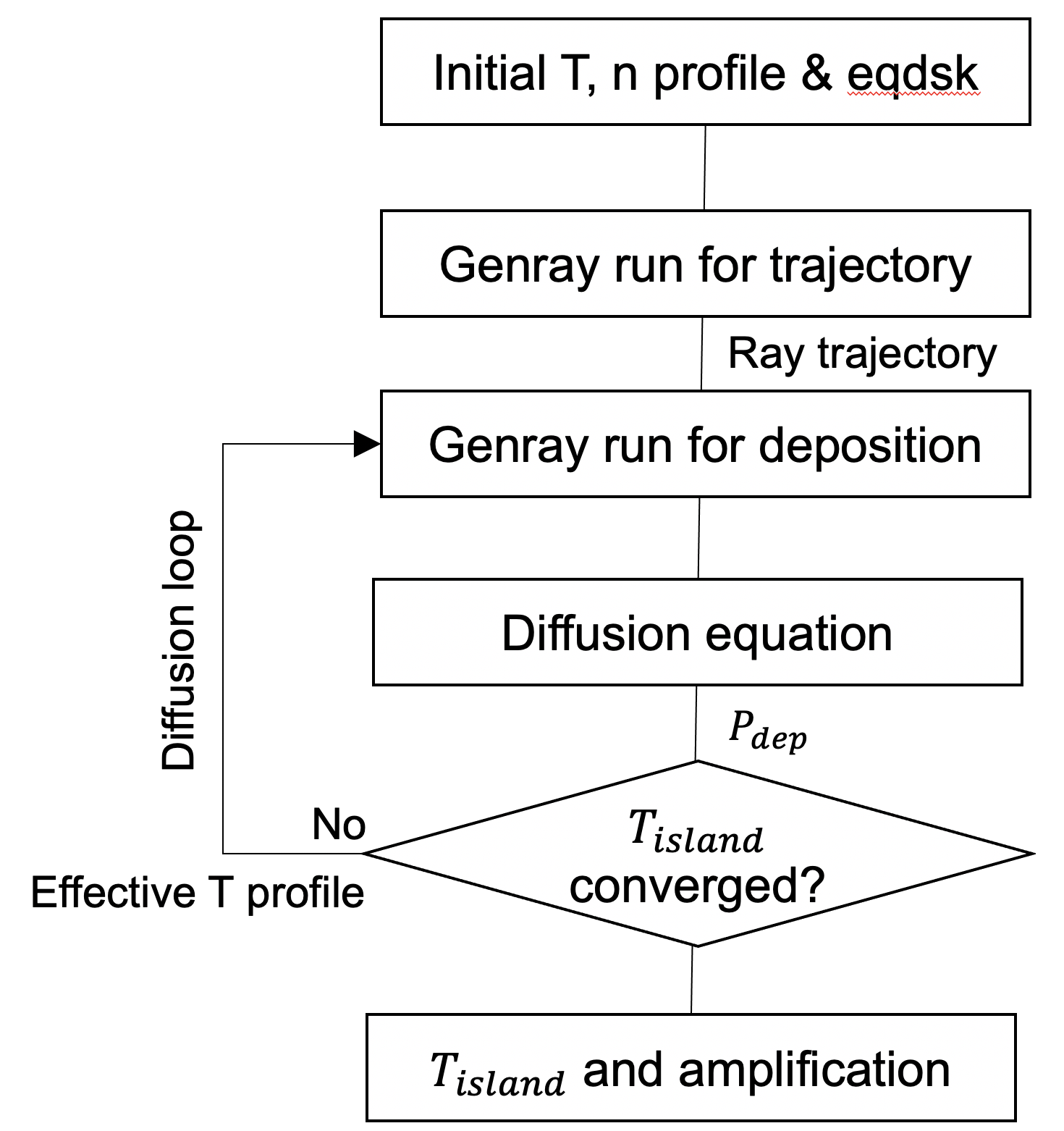}
\caption{\label{fig:intro_flow} OCCAMI flow chart.}
\end{figure}

The flow of the logic in the OCCAMI code is shown in Fig. (\ref{fig:intro_flow}) During initialization, GENRAY
calculates the ray trajectory in the initial axisymmetric equilibrium. The temperature and density profile along the trajectory are updated using the 3D perturbed flux surfaces, with flattened density and temperature in the island region. The ray  trajectory and the power deposition along the trajectory are then recalculated by GENRAY using the modified density and temperature along the trajectory. The small resonant magnetic field perturbation producing the island is neglected during ray tracing. The power deposited in the interior of a given flux surface in the island is calculated by subtracting the power in the wave when it exits that region from the power in the wave when it enters that region.

It is assumed that the temperature is constant on the flux surfaces in the island. At each iteration, the code  updates the temperature profile in the island by solving the cross-field heat diffusion equation with the calculated power deposition as the source term. The diffusion equation can be analytically integrated once to reduce its order, yielding the first order ODE
\begin{equation}\label{eq_diff}
\frac{d u}{d \sigma} = -\frac{P_\text{dep}(\sigma)}{n\kappa_{\perp}T_s}\frac{\sigma}{E(\sigma)-(1-\sigma^2)K(\sigma)}\frac{WM}{32\pi r_r R_0},
\end{equation}
where $u=(T-T_s)/T_s$ is the normalized island temperature; $T_s$ is the separatrix temperature; $\sigma$ is an island radial flux surface coordinate that ranges from 0 at the O-point to 1 at the separatrix; $P_\text{dep}$ is the power deposition; $n$ is the density; $E$ and $K$ are are complete elliptical integrals of the first and second kind; $W$ is a prescribed island width; $M$ is the island poloidal mode number; $R_0$ is the major radius. $r_r$ is the minor radius at the rational surface. The updated temperature profile is fed back to GENRAY to obtain  the $P_\text{dep}$ profile for use in the next iteration. The solution is considered to be converged when the change in the island temperature between successive iterations is sufficiently small.

The solution for the temperature obtained after the first iteration corresponds to the conventional linear solution. In the following section, we will denote the value of $u$ at the O-point by $u_0$. We define an amplification factor by the ratio of the the converged nonlinear value of $u_0$ to its linear value, $A \equiv u_0/u_0^{lin}$. This will provide a measure of the strength of the nonlinear effect. 

\section{\label{sec:diagnostics} Diagnostic Considerations}

In an experiment, our goal is to measure the temperature perturbation in the magnetic island. We plan to use ECE for this purpose, which has been shown to provide well resolved measurements of temperature perturbations in magnetic islands. \cite{bardoczi_diffusivity_2016,bardoczi_diffusivity_2017} There is a complication for shot 141060 that the field is sufficiently low that the ECE emited at  the 2nd harmonic surface encounters the third harmonic surface before exiting the plasma. Some of the power is absorbed at the third harmonic surface, and emission from the third harmonic surface pollutes the signal. A method has previously been developed to correct for this using a 1D radiation transport model. \cite{austin1996electron} For the purpose of measuring the temperature in a rotating island, there is an advantage that the ECE signal produced by the temperature perturbation in the island is time dependent, while the ECE emitted from the third harmonic surface is stationary. This allows the development of an improved method for reconstructing the temperature perturbation in a rotating island. \cite{Jung}

In shot 141060, the island locked at about $1750$ms. The above discussion suggests that it is advantageous for diagnostic purposes to maintain the rotation of the island. This can be done by entraining the island in a rotating RMP. Choi \textit{et al} \cite{choi_feedforward_2018} describe how such an island can be entrained at 70 Hz using feed forward entrainment. 

We will see below that the nonlinear response of rotating islands to ECCD is reduced, relative to the response of locked islands. The broader power deposition in rotating islands causes the outermost flux surfaces in the island to more rapidly exceed the microinstability threshold, with the boundary of the region where the threshold is exceeded gradually moving in towards the center of the island. The EC can be modulated to deal with this. As described in Volpe \textit{et al} 2015\cite{volpe_2015}, and Choi \textit{et al} 2018\cite{choi_feedforward_2018}, the entrainment of the island aids in accurately maintaining the optimal phase of the modulation.

\begin{figure}
\centering
\includegraphics[width=0.49\textwidth]{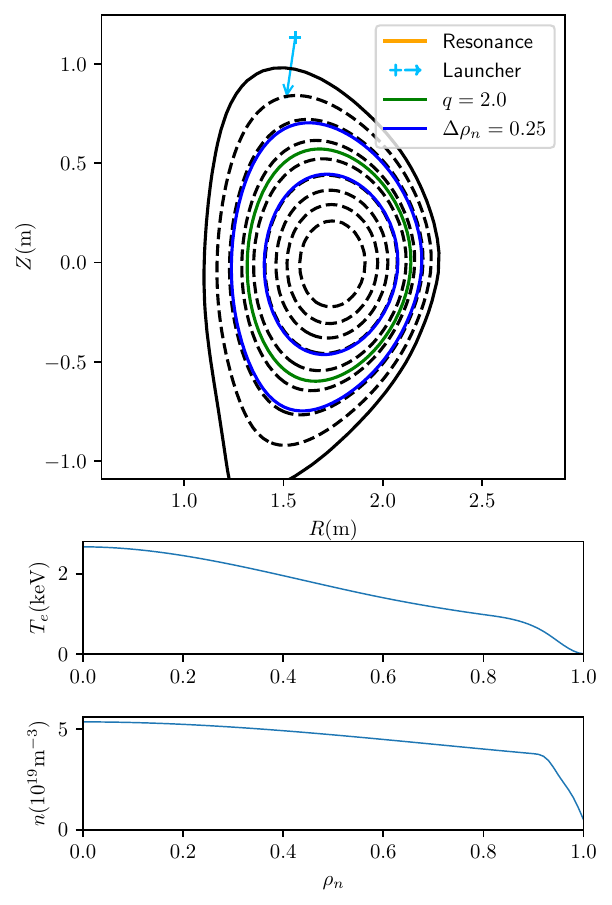}
\caption{\label{fig:1735eq} EFIT reconstructed equilibrium and profiles for shot DIII-D 141060 at $t=1735$ms}
\end{figure}

\begin{figure*}
     \centering
     \begin{subfigure}[c]{0.49\textwidth}
         \centering
         \includegraphics[width=\textwidth]{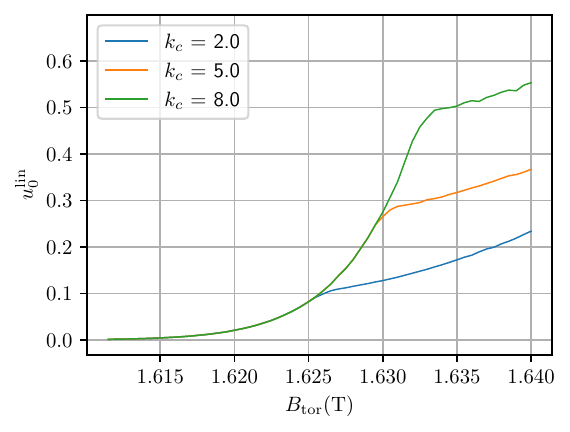}
         \caption{Locked $25\%$ island, linear normalized island O-point temperature perturbation $u_0^{lin}$} vs. $B_\text{tor}$.
         \label{fig:result_shifted_25_locked_u0lin}
     \end{subfigure}
     \hfill
     \begin{subfigure}[c]{0.49\textwidth}
         \centering
         \includegraphics[width=\textwidth]{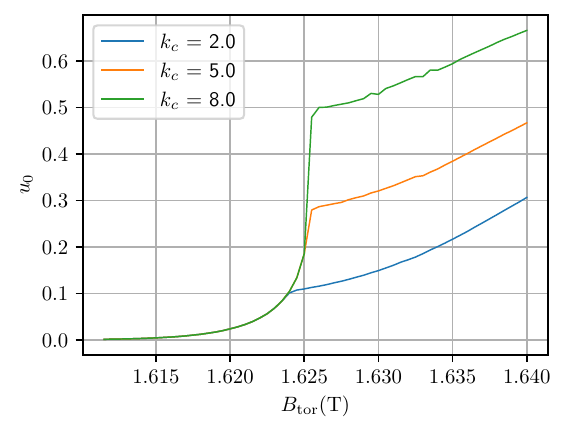}
         \caption{Locked $25\%$ island, nonlinear normalized island O-point temperature perturbation $u_0$ vs. $B_\text{tor}$.}
         \label{fig:result_shifted_25_locked_u0}
     \end{subfigure}
    
     \centering
     \begin{subfigure}[c]{0.49\textwidth}
         \centering
         \includegraphics[width=\textwidth]{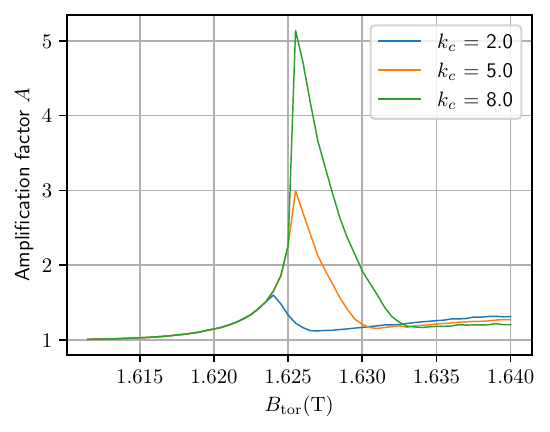}
         \caption{Locked $25\%$ island, amplification $A$ vs. $B_\text{tor}$}
         \label{fig:result_shifted_25_locked_A}
     \end{subfigure}
     \hfill
     \begin{subfigure}[c]{0.49\textwidth}
         \centering
         \includegraphics[width=\textwidth]{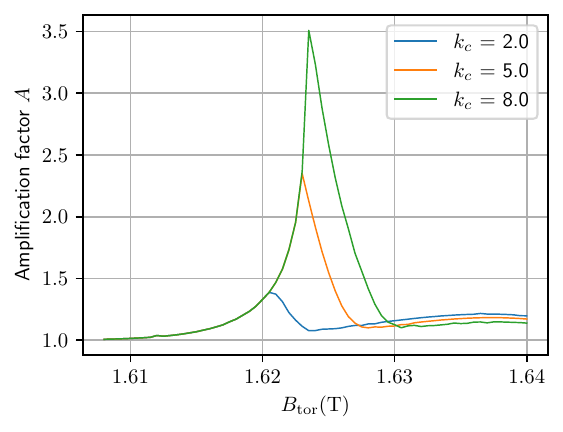}
         \caption{Locked $20\%$ island, amplification $A$ vs. $B_\text{tor}$}
         \label{fig:result_shifted_20_locked_A}
     \end{subfigure}
     
     \centering
     \begin{subfigure}[c]{0.49\textwidth}
         \centering
         \includegraphics[width=\textwidth]{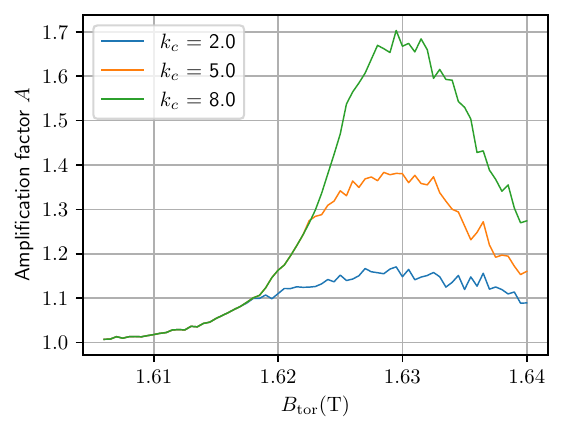}
         \caption{Rotating $25\%$ island, amplification $A$ vs. $B_\text{tor}$}
         \label{fig:result_25_rot_A}
     \end{subfigure}
     \hfill
     \begin{subfigure}[c]{0.49\textwidth}
         \centering
         \includegraphics[width=\textwidth]{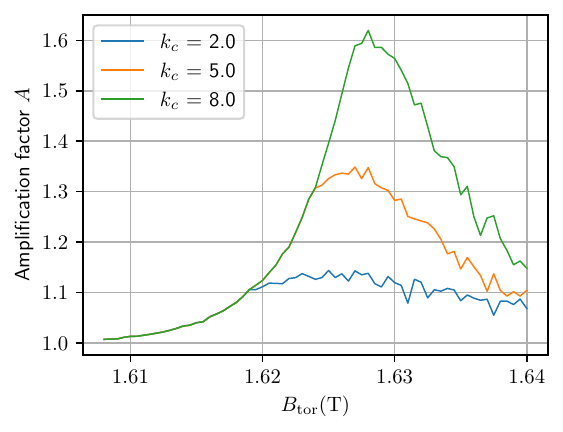}
         \caption{Rotating $20\%$ island, amplification $A$ vs. $B_\text{tor}$}
         \label{fig:result_20_rot_A}
     \end{subfigure}
        \caption{$B_\text{tor}$ scan results for DIII-D shot 141060 EFIT reconstructed equilibrium with $\kappa_s/\kappa_0 = 3.0$, 1 MW of top launch EC power.}
        \label{fig:result_shifted_rot_25_locked_A}
\end{figure*}

\section{\label{sec:results} Calculations for DIII-D}

\subsection{\label{sec:parameters}Simulation setup for DIII-D}

For the DIII-D reference equilibrium, we use the reconstructed equilibrium for DIII-D shot 141060 at $1735$ms. The plasma cross-section and the density and temperature profiles are shown in Fig, \ref{fig:1735eq}. The separatrix temperature $T_s=1.235$keV. During the shot, the islands lock shortly after$t=1735$ms. This allows us to use the same equilibrium for both the rotating and locked islands, and it provides a comparison of the nonlinear effects in a locked island with those in a rotating island. The shot was part of the 2015 DIII-D experiment studying RF stabilization of locked islands.
\cite{volpe_2015} In this paper, we consider islands having widths of $20\%$ and $25\%$ of the minor radius.

We initialize our ray trajectories at the locations of the top launchers installed in DIII-D, $R = 1.559m, Z = 1.133m, \phi = 90\degree$ and $300\degree$. The toroidal and poloidal launch angles of the top launchers are,  $\alpha = 241\degree$ ($0\degree$ points to $+\hat{R}$ and $90\degree$ points to $+\hat{\phi}$), and $\beta=162\degree$ ($0\degree$ points to $+\hat{Z}$ and $90\degree$ points to $-\hat{R}$).\cite{chen_doubling_2022} The launchers were installed with fixed launch angles. In the absence of the ability to steer the launch angles, we adjust the strength of the toroidal field $B_\text{tor}$ to control the deposition profiles. This shifts the location of the second harmonic resonance, and thus the location of the power deposition. (Second harmonic X-mode heating is used, with the launchers operating at $110$GHz.)

For the simulations, as $B_\text{tor}$ was modified, the current density was correspondingly modified to approximately preserve the $q$ profile. The temperature and density profiles were not modified. EFIT was used to recalculate the equilibrium for each new value of $B_\text{tor}$ for which a simulation was performed. 

With the toroidal launch angle fixed, the direction of the toroidal field matters. Our simulations use toroidal field direction $B_\text{tor}>0$ (counterclockwise when viewed from above), for which the launch angles were optimized.

\subsection{\label{sec:Bt_sweeps} Sweeps in $B_\text{tor}$}

For the experiment, we propose to scan over a range of values of $B_\text{tor}$ by continuously varying the current in the toroidal field coil, and observe the island temperature perturbation, primarily with ECE diagnostics. This would be done for several different magnitudes of the EC power. At relatively low power, we expect the dependence of the temperature perturbation on the power to be linear, except when the microinstability threshold is encountered. When the temperature gradient is below the threshold, the non-linearity in the temperature response measures the strength of RF condensation.

We use the OCCAMI code to calculate the predicted temperature perturbation as a function of $B_\text{tor}$. Figure \ref{fig:result_shifted_rot_25_locked_A} shows a set of $B_\text{tor}$ scans with $1$MW of injected EC power. Figures \ref{fig:result_shifted_25_locked_u0lin} to \ref{fig:result_shifted_25_locked_A} show the results of calculations for a locked island whose width is $25\%$ of the minor radius. Here $u_0$ is the value of $u=(T-T_s)/T_s$ at the island O-point. Figure \ref{fig:result_shifted_25_locked_u0lin} shows the linear solution for $u_0$ vs. $B_\text{tor}$, while figure \ref{fig:result_shifted_25_locked_u0} shows the nonlinear solution. Every curve in each plot correspond to a different value of the microinstability threshold parameter. The curves are identical below the thresholds. The power deposited in the island increases as $B_\text{tor}$ increases. When the power deposition is small, the effect of the nonlinearity is small. As the power deposition increases, the nonlinear solution increases more rapidly. There is a discontinuous decrease in the slope when the microinstability threshold is encountered. 

Fig. \ref{fig:result_shifted_25_locked_A}  shows the amplification factor $A \equiv u_0 / u_0^{lin}$ for the $25\%$ island. There is a narrow region just below the ITG threshold where the amplification is rapidly increasing.

In practice, we expect to see a transition from the nonlinear $B_\text{tor}$ dependence of Fig. \ref{fig:result_shifted_25_locked_u0} to the linear dependence of \ref{fig:result_shifted_25_locked_u0lin} as we decrease the EC input power. (We linearly rescale the lower power measurements for comparison with the higher power.)  The range of $B_\text{tor}$ just below the ITG threshold will be of greatest interest. After seeing the $u_0$ dependence shown in \ref{fig:result_shifted_25_locked_u0}, we will want to narrow the range of $B_\text{tor}$ in subsequent shots to examine this region more closely. 

Figure \ref{fig:result_shifted_20_locked_A} shows the calculated amplification for a locked $20\%$ island. The magnitude of the amplification is smaller than that for a $25\%$ island, but the shape of the amplification vs. $B_\text{tor}$ curve is approximately the same. 

Fig. \ref{fig:result_25_rot_A} shows the amplification curve for a rotating island whose width is $25\%$ of the minor radius. The decrease in the amplification is accounted for to some extent by the fact that more of the power is now being deposited outside the island. However the shape of the curve is quite different from that for the locked islands. This is explained by the fact that a larger fraction of the power is deposited near the periphery of the island. The temperature gradient on the peripheral island flux surfaces encounters the microinstability threshold before the flux surfaces that are closer to the center. The island enters the stiff regime more gradually, with the boundary of the region where the flux surfaces encounter the threshold moving gradually inward.

Fig. \ref{fig:result_shifted_rot_25_locked_A} shows the amplification curve for a rotating $20\%$ island. The magnitude of the amplification is smaller than that for a $25\%$ island, but the shape of the curve is approximately the same.

\begin{figure}
\includegraphics[width=0.49\textwidth]{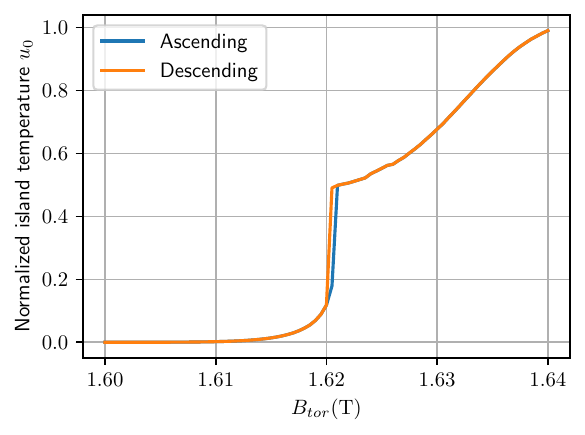}
\caption{Hysteresis in $25\%$ island with 3MW of EC power, $k_c=8$, $\kappa_s/\kappa_0 = 3.0$}
\label{fig:result_25_locked_hys}
\end{figure}


\subsection{\label{sec:bifurcation} Bifurcation and hysteresis}

Under suitable circumstances, nonlinear effects from RF condensation can produce a bifurcation of the solution to the nonlinear steady-state thermal diffusion equation. This bifurcation has two diagnostic signatures: a discontinuous jump in the temperature, or a hysteresis when a launcher parameter continuously rises above, and then returns below the discontinuity threshold. Experimentally, a sufficiently strong hysteresis can serve as an alternative signature to the nonlinear temperature growth discussed in Section\ref{sec:Bt_sweeps}. Numerically, hysteresis is the preferable evidence for bifurcation, because it is impossible to distinguish between a rapid increase in $u_0$ with a true discontinuity due to the finite resolution of configuration scans.

We demonstrate the possibility to experimentally produce a temperature bifurcation at DIII-D by simulating a hysteresis due to RF condensation. Fig. \ref{fig:result_25_locked_hys} shows a hysteresis curve for a $25\%$ island in shot 141060 with $3MW$ of injected power. As $B_\text{tor}$ increases, there is a threshold point at approximately $1.62 T$ above which $u_0$ increases rapidly before encountering a discontinuous slope and a slower increase. Starting at that solution and decreasing $B_\text{tor}$, the solution now goes past the previously encountered threshold value of $B_\text{tor}$ for a short distance before dropping back to the previously encountered solution. Empirically, the uncertainty in toroidal magnetic field control is $0.05T$. The hysteresis is likely too weak to observe experimentally. However, it is still of theoretical interest, because it confirms that the jump in $u_0$ in this configuration is indeed a true discontinuity.

\section{Analysis of a DIII-D $B_\mathrm{tor}$ scan \label{analysis}}

\begin{figure*}\centering
     \begin{subfigure}[b]{0.3\textwidth}
         \centering
         \includegraphics[width=\textwidth]{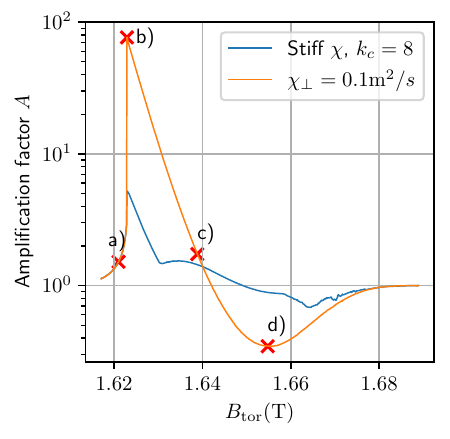}
     \end{subfigure}
     \hfill
     \begin{subfigure}[b]{0.3\textwidth}
         \centering
         \includegraphics[width=\textwidth]{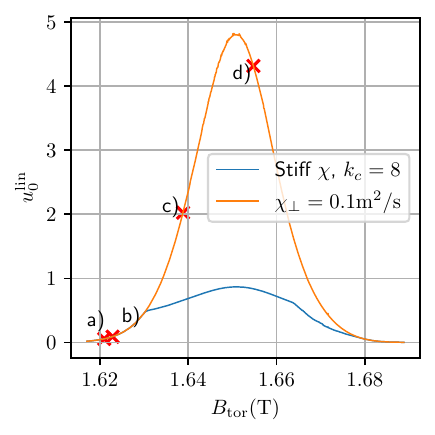}
     \end{subfigure}
     \hfill
     \begin{subfigure}[b]{0.3\textwidth}
         \centering
         \includegraphics[width=\textwidth]{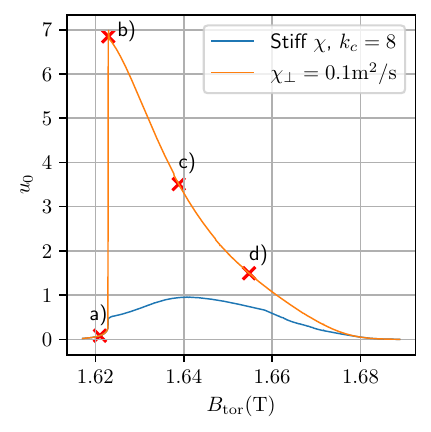}
     \end{subfigure}
    \caption{Stages in a $B_\text{tor}$ sweep, run with locked $25\%$ islands and $P = 1MW$. Compare the constant and stiff $\kappa_\perp$ traces. Note that stiffness threshold limits island heating and saturates the non-linear effect.}
    \label{fig:discussion_stages}
\end{figure*}

We take a closer look at the $B_\mathrm{tor}$ scan shown in Figs. \ref{fig:result_shifted_25_locked_u0lin} - \ref{fig:result_shifted_25_locked_A} for a $25\%$ locked island with $1$MW of injected EC power. Figure \ref{fig:discussion_stages} shows plots of the amplification factor, $A$, the linear normalized temperature perturbation in the island, $u_0^{lin}$, and the nonlinear normalized temperature perturbation in the island, $u_0$, as a function of $B_\text{tor}$, for both the constant $\kappa_\perp$ and stiff diffusivity models. Point (b) corresponds to the value of $B_\text{tor}$ where the constant $\kappa_\perp$ model sees the largest amplification. Point (a) corresponds to a slightly lower value of $B_\text{tor}$. Point (d) corresponds to the value of $B_\text{tor}$ where the constant $\kappa_\perp$ model sees the lowest amplification. Point (c) corresponds to a value of $B_\text{tor}$ intermediate between points (b) and (d). The values of $|B_\text{tor}|$ at the center of the plasma for the four points (a), (b), (c) and (d) are, respectively,
1.621 T, 1.623 T, 1.639 T and 1.655 T.

\begin{figure*}
    \centering
    \includegraphics[width=\textwidth]{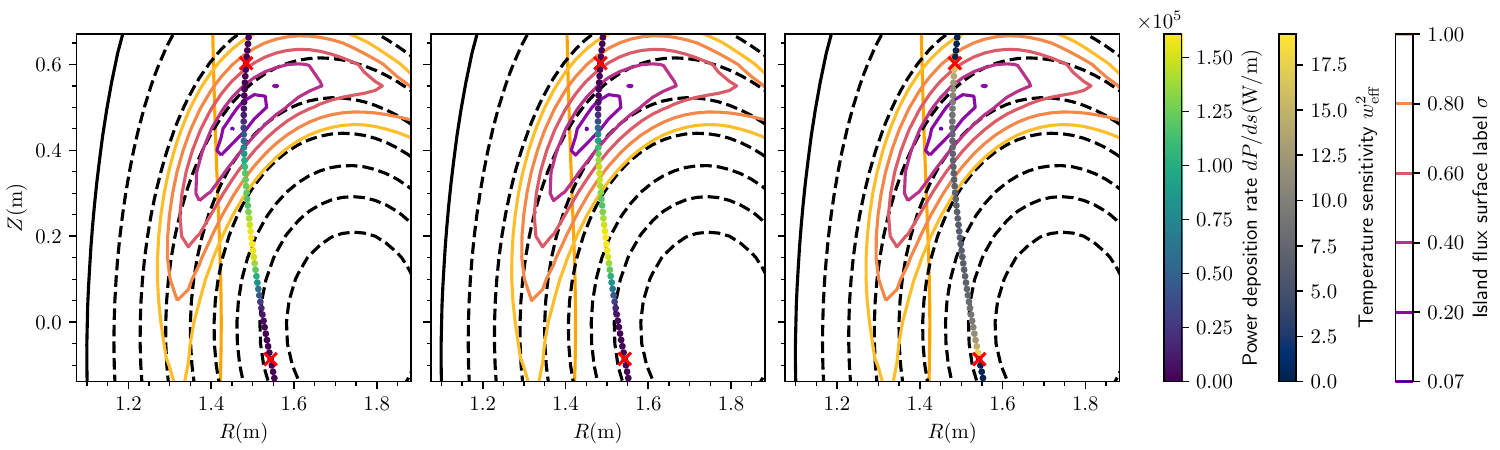}
    \caption{From left to right, the $(dP/ds)_\text{lin}$, amplified $dP/ds$, and $w_\text{eff}^2$ in the rising stage of a $B_\text{tor}$ scan. (point (a) in Fig. \ref{fig:discussion_stages})}.
    \label{fig:app_stages_rising}
\end{figure*}

\begin{figure*}
    \centering
    \includegraphics[width=\textwidth]{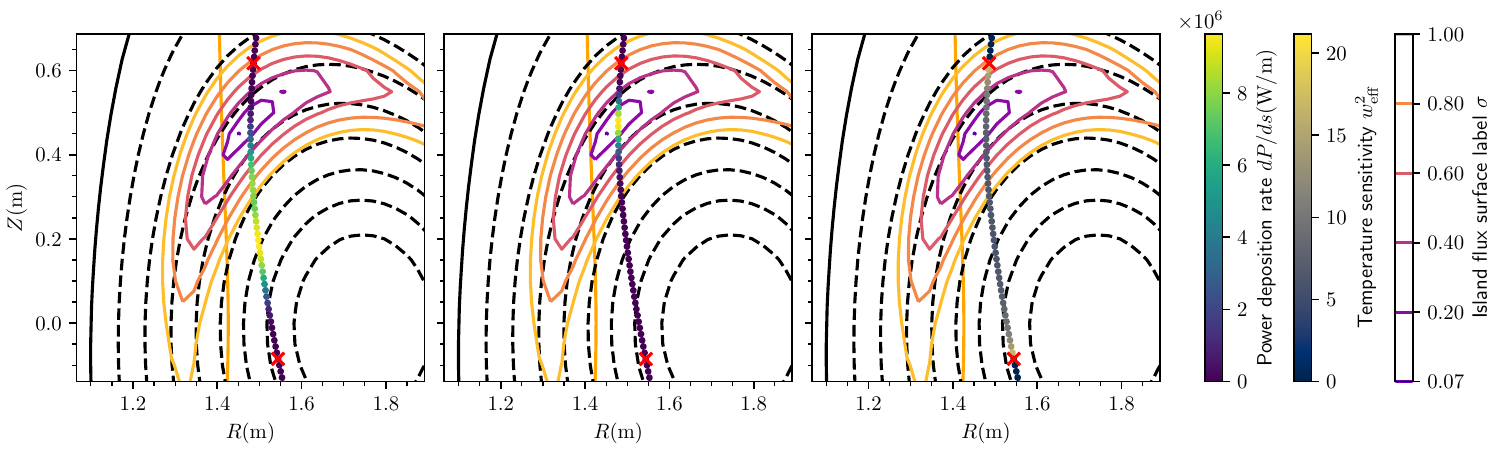}
    \caption{From left to right, the $(dP/ds)_\text{lin}$, amplified $dP/ds$, and $w_\text{eff}^2$ at the $A$ maximum of a $B_\text{tor}$ scan. (point (b) in Fig. \ref{fig:discussion_stages})}.
    \label{fig:app_stages_max_amp}
\end{figure*}

\begin{figure*}
    \centering
    \includegraphics[width=\textwidth]{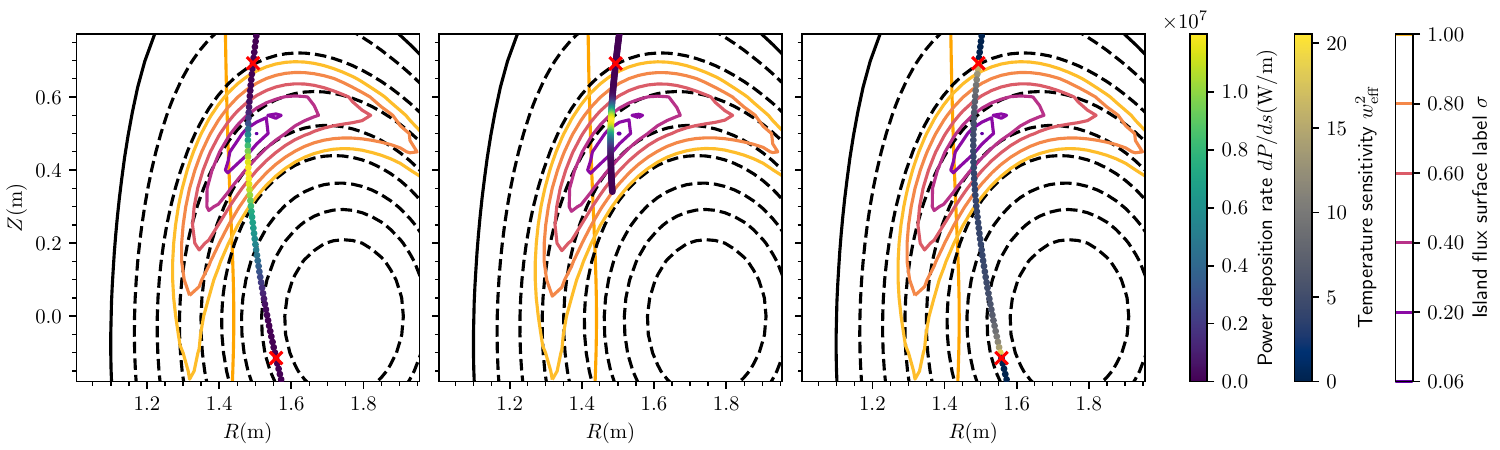}
    \caption{From left to right, the $(dP/ds)_\text{lin}$, amplified $dP/ds$, and $w_\text{eff}^2$ in the falling stage of a $B_\text{tor}$ scan. Note that the ray's $dP/ds$ maximum occurs prior to reaching the O-point (point (c) in Fig. \ref{fig:discussion_stages})}.
    \label{fig:app_stages_falling}
\end{figure*}

\begin{figure*}
    \centering
    \includegraphics[width=\textwidth]{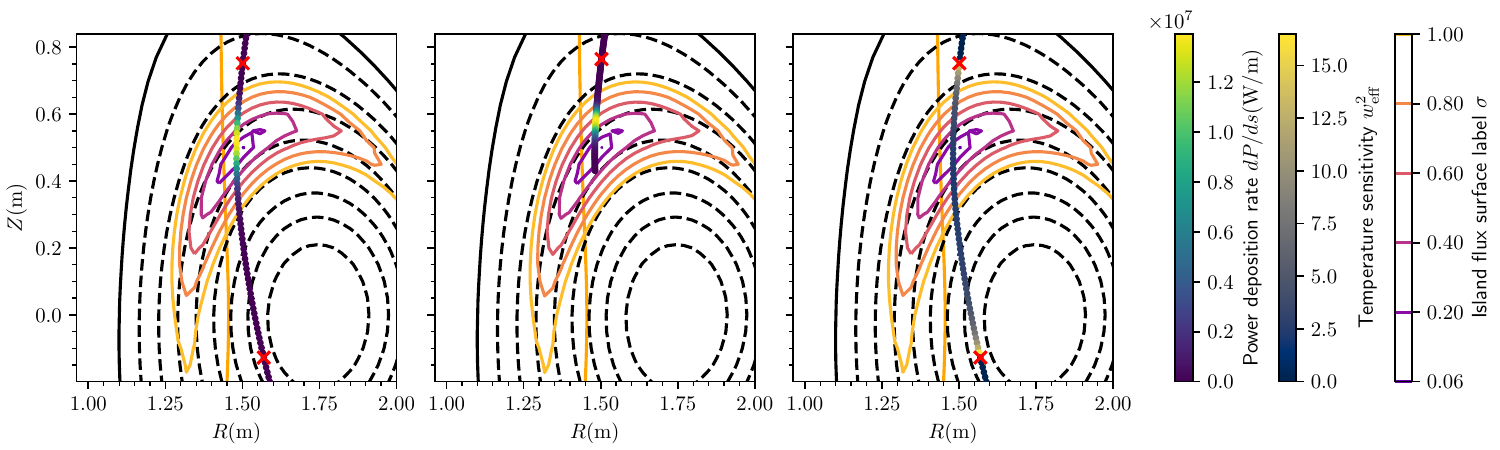}
    \caption{From left to right, the $(dP/ds)_\text{lin}$, amplified $dP/ds$, and $w_\text{eff}^2$ at the $A$ minimum of a $B_\text{tor}$ scan. Note the strong shadowing effect. (point (d) in Fig. \ref{fig:discussion_stages})}.
    \label{fig:app_stages_min_amp}
\end{figure*}

Figure \ref{fig:appendix_weff} shows $w_\text{eff}^2$ along the trajectories corresponding to each of the four points. Figures \ref{fig:app_stages_rising} to \ref{fig:app_stages_min_amp} show the plasma cross-section with $w_\text{eff}^2$, the linear power deposition, $dP/ds_{lin}$, and the nonlinear power deposition, $dP/ds$, along the trajectories for each of the four points.


As $B_\text{tor}$ increases, the EC resonance layer moves to larger $R$, closer to the ray trajectory. As a result, the peak of $P^{lin}_\text{dep}$, the power deposition predicted by the linear theory, moves earlier along the trajectory. At stage $(a)$ in Fig. \ref{fig:discussion_stages}, the strength of the non-linear effect first gradually increases when the $P^{lin}_\text{dep}$ tail enters the island. In Fig. \ref{fig:app_stages_rising}, we can see that the nonlinear effect is small at this stage. At stage $(b)$, When $P^{lin}_\text{dep}$ at the O-point is sufficiently large, the nonlinear effect is strong, and $u_0$ increases rapidly with increasing $B_\text{tor}$. There is a large nonlinear shift in the position of the deposition profile along the ray trajectory, as shown in Fig. \ref{fig:app_stages_max_amp}. Note that the effect of the microinstability threshold for the stiff diffusivity model causes the island temperature perturbation to saturate. The extent to which the power deposition nonlinearity affects $u_0$ is affected by the microinstability threshold in the island, which is unknown at present. The bulk of $P_\text{dep}$ moves from the resonance to island o-point. In stage $(c)$, the further increase in $B_\text{tor}$ causes the deposition peak to move into the region past the O-point, as shown in Fig. \ref{fig:app_stages_falling}. We are now beginning to see the shadowing effect. In stage $(d)$, shown in Fig. \ref{fig:app_stages_min_amp}, the shadowing effect reduces the steady-state island temperature $u_0$ below its linear model prediction $u_0^{lin}$ and results in $A<1$.

In addition to the amplification of $u_0$ relative to $u_0^{lin}$, the asymmetry of the $u_0$ curve compared to the $u_0^{lin}$ curve seen in Fig. \ref{fig:discussion_stages} may serve as an additional experimental signature of the nonlinear effect. However, the extent to which the asymmetry is present can be strongly affected by the microinstability threshold.

\section{\label{sec:AUG} Calculations for ASDEX Upgrade (AUG)} \label{additional}

\begin{figure*}[h]
     \centering
     \begin{subfigure}[b]{0.49\textwidth}
         \centering
         \includegraphics[width=\textwidth]{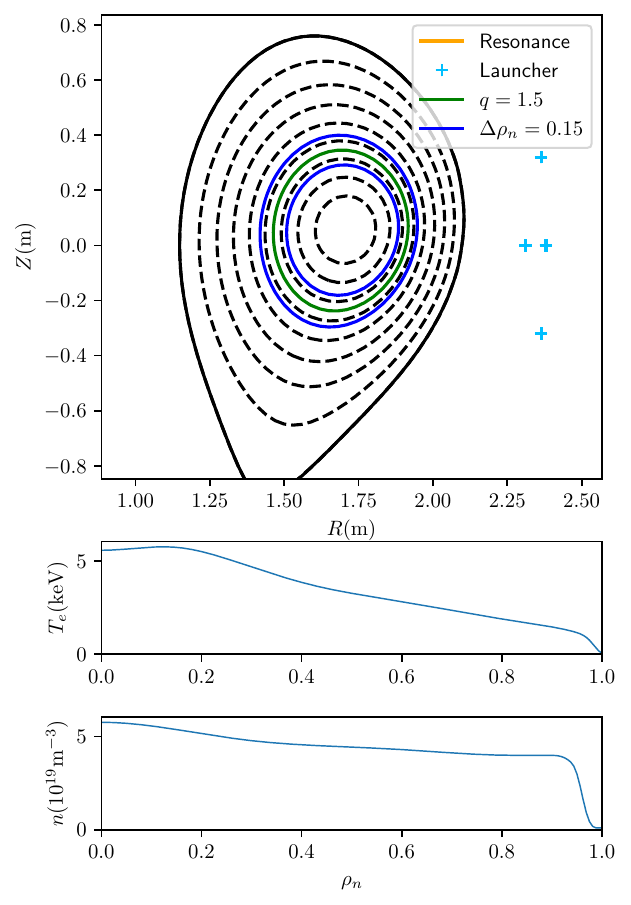}
         \caption{$t=3400ms, q=3/2, 15\%$ island}
         \label{fig:eq_aug_34}
     \end{subfigure}
     \hfill
     \begin{subfigure}[b]{0.49\textwidth}
         \centering
         \includegraphics[width=\textwidth]{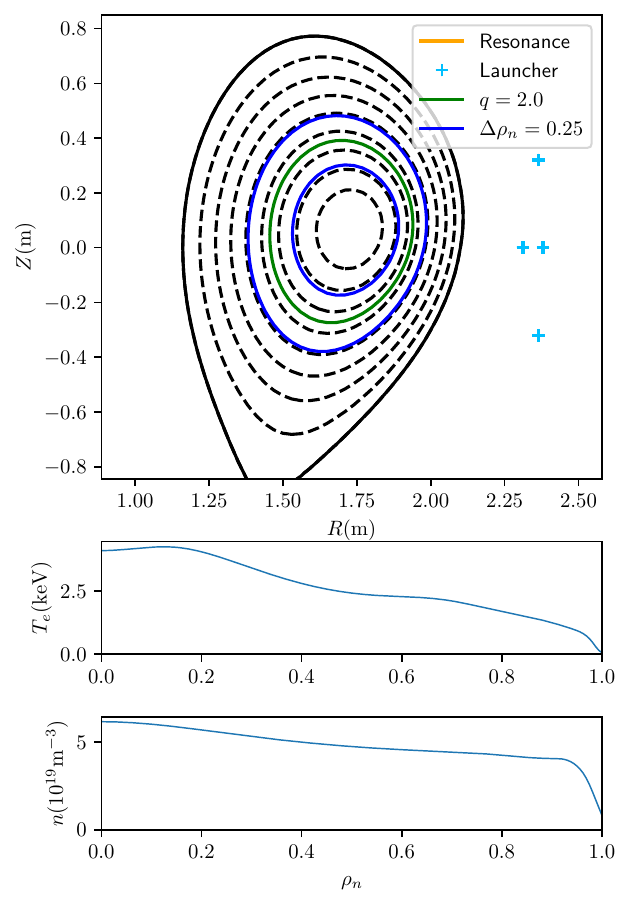}
         \caption{$t=5400ms, q=2, 25\%$ island}
         \label{fig:eq_aug_54}
     \end{subfigure}
        \caption{Equilibrium and profiles for AUG shot 35350, at two chosen times.}
        \label{fig:eq_aug}
\end{figure*}

An AUG equilibrium was used to study nonlinear effects for outside launch. The calculations considered the stabilization of rotating $q=3/2, 15\%$ and $q=2, 25\%$ islands at 3400 and 5400 ms of AUG shot 35350. The profiles and plasma cross-section at these two times are shown in Fig. \ref{fig:eq_aug}. The solid blue lines indicate the edges of the regions subtended by the islands, and the solid green lines indicate the locations of the rational surfaces. The red "$+$" signs indicate the locations of the EC launchers. The vertical orange lines indicate the position of the second harmonic EC resonance.

AUG has 8 EC launchers on the low field side of the plasma, located at major radii $R$ ranging from $R=2.312$ to $R=2.38$, at $z = \pm 0.32$ and $z=0$, and at 4 different azimuthal angles, $\phi$\cite{lohr_electron_2005}. Each AUG launcher has a steerable mirror with sufficient range of the poloidal launch angle to cover the entire core plasma, and with the range of the toroidal launch angle of $\alpha = 180 \degree \pm 25\degree$. This range of the toroidal launch angle is sufficient to cover the region where the current drive efficiency is maximized. The range of $\alpha$ is to be contrasted with the toroidal launch angle for the DIII-D top launch experiments, which is $61 \degree$ relative to the $-\hat{R}$ direction.

\begin{figure}[!ht]
    \centering
    \begin{subfigure}[b]{0.49\textwidth}
        \centering
        \includegraphics[width=\textwidth]{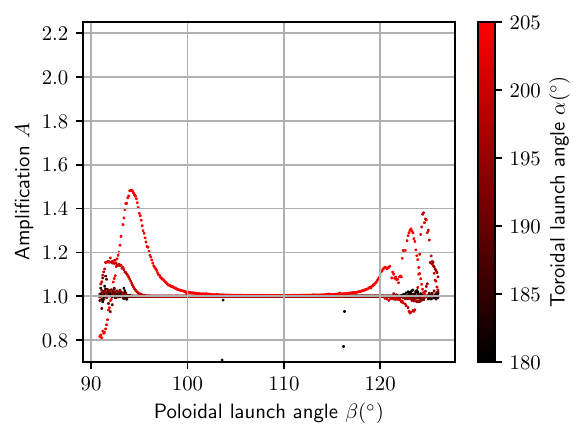}
        \caption{$t=3400$ms}
        \label{fig:result_aug_34}
    \end{subfigure}
    \hfill
    \begin{subfigure}[b]{0.49\textwidth}
        \centering
        \includegraphics[width=\textwidth]{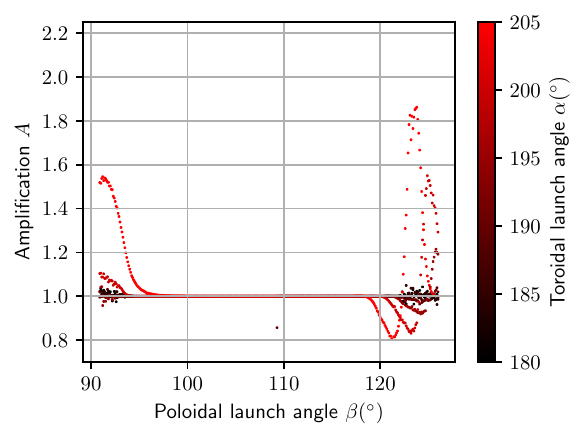}
        \caption{$t=5400$ms}
        \label{fig:result_aug_54}
    \end{subfigure}
    \caption{Amplification $A$ with $1 MW$ EC power in AUG shot 35350}
    \label{fig:result_aug}
\end{figure}

\begin{figure}[!ht]
    \centering
    \begin{subfigure}[b]{0.49\textwidth}
        \centering
        \includegraphics[width=\textwidth]{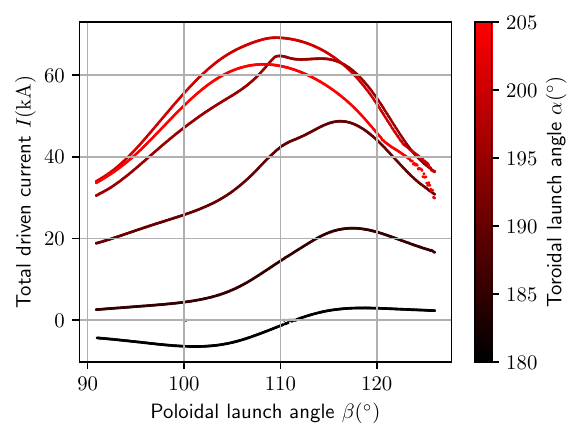}
        \caption{$t=3400$ms}
        \label{fig:result_aug_34_I}
    \end{subfigure}
    \hfill
    \begin{subfigure}[b]{0.49\textwidth}
        \centering
        \includegraphics[width=\textwidth]{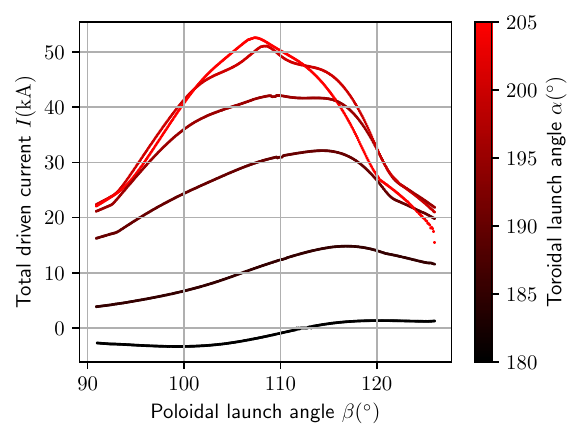}
        \caption{$t=5400$ms}
        \label{fig:result_aug_54_I}
    \end{subfigure}
    \caption{Total EC driven current $I$ with $1MW$ EC power in AUG shot 35350.}
    \label{fig:result_aug_I}
\end{figure}

The islands were taken to be rotating for the AUG calculations reported here. For simplicity, and to accelerate the calculations, the AUG calculations assume that there is a single ray originating from the launcher at $R=2.364$, $z=0.32$, $\phi = 98.46\degree$. The diffusivity was taken to be a constant, with $\kappa_\perp =0.1 \text{m}^2/\text{s}$. The injected EC power was taken to be $1$MW. The calculated driven EC current is shown as a function of the poloidal launch angle for several different values of the toroidal launch angle in Fig. \ref{fig:result_aug_I}. The current drive efficiency peaks at $\alpha \approx 180\degree\pm20\degree$.

Fig. \ref{fig:result_aug} shows the calculated amplification as a function of the poloidal launch angle for several different values of the toroidal launch angle. Values of amplification up to $A\approx1.8$ are seen. The highest amplification is seen at the toroidal launch angles having the largest distance from $\alpha = 180 \degree$, and for the launchers that are off the midplane. 
It can be seen in Figures \ref{fig:result_aug} that the amplification is increasing very rapidly as the largest toroidal launch angles are approached.



\section{\label{sec:discussions} Conditions governing the magnitude of the nonlinear effects}


Significant non-linear effects can be seen when the power deposition and consequent heating of the plasma along an EC ray trajectory causes the power deposition profile to shift earlier along the trajectory. When the ray trajectory passes through a large, heated island, there is a local maximum in the temperature profile in the interior of the island. This can lead to coalescence of the power deposition near the temperature maximum. The shift of the power deposition location is constrained by the relativistic constraint discussed in Section \ref{sec:power}, which limits the region where the EC power may be deposited. RF condensation is most readily seen when damping is permitted at the local temperature maximum in the island interior, and in a significant portion of the ray trajectory lying in the island region beyond the temperature maximum. The nonlinearity can also lead to a shadowing effect, in which the power is deposited before the temperature maximum is reached. The relativistic constraint can be helpful here if it limits the region where the power can be deposited before the temperature maximum is reached.

\begin{figure*}
    \centering
    \begin{subfigure}[c]{0.6\textwidth}
        \centering
        \includegraphics[width=\textwidth]{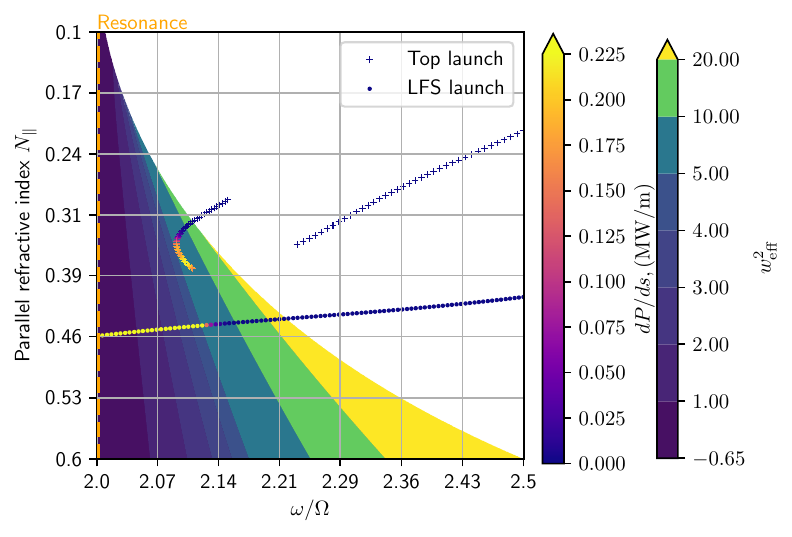}
        \caption{DIII-D $w_{eff}^2(T_0, N_{\parallel}, \omega/\Omega)$, with in-island ray trajectory and deposition rate. $N_{\parallel}$ depends on the ray's alignment with magnetic field. $\omega/\Omega \propto R$ through $\Omega$. $T_0$ is the unmodified, unperturbed island temperature. The left edge of the plot, $\omega/\Omega = 2$, is the resonance.}
        \label{fig:discussion_top_side_weff}
    \end{subfigure}
    \hfill
    \begin{subfigure}[c]{0.3\textwidth}
        \centering        
        \includegraphics[width=\textwidth]{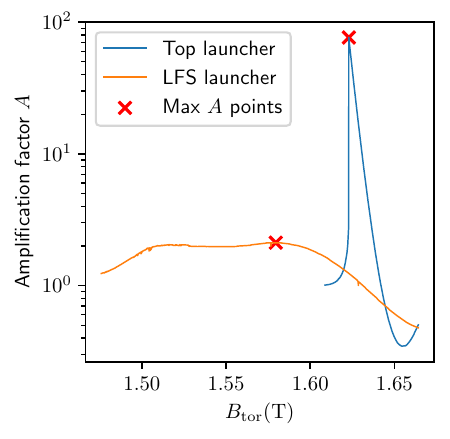}
        \caption{Comparison of $A$ between top and LFS launch, calculated with constant $\chi_{\perp}$. Max $A$ cases shown in left are highlighted.}
        \label{fig:discussion_top_side_A}
    \end{subfigure}
    \caption{Comparison of $w_{eff}^2$, in-island trajectories and deposition rate(left) at max $A$ cases (highlighted in right) for top and low field side launchers. Note that the ray from the top launcher does not damp completely, and crosses the island chain again. The simulation is run with constant $\kappa_{\perp}=0.1\text{m}^2/\text{s}$ and $P=1$MW. }
    \label{fig:discussion_top_side_both}
\end{figure*}

As discussed in Section \ref{sec:power}, the strength of the nonlinear effects also depends on the value of $w_\text{eff}$. Figure \ref{fig:discussion_top_side_weff} shows a top launch ray trajectory and an outside launched ray trajectory in DIII-D as a function of $\omega/\Omega$ and $N_{\parallel}$, with the color coding indicating the values of $w_\text{eff}^2$ and $dP/ds$. Here $w_\text{eff}^2$ has been evaluated using an approximate analytic formula, as discussed in Appendix \ref{app:weff_goodness}. The trajectories correspond to the maximum amplification points in two $B_\text{tor}$ scans, as indicated in Fig. \ref{fig:discussion_top_side_A}. Only the portions of the trajectories that lie in the island are shown in the figure. The top launch ray trajectory is the same as that shown in Fig. \ref{fig:intro_example}. That trajectory turns around in the island interior, exits the island and then crosses the island chain again. 

For $\omega/\Omega>2$, $w_\text{eff}^2$ monotonically increases with $\omega/\Omega \propto R$. Relative to the trajectory launched from the low field side, the top launch trajectory spends more of its time in a region farther from the second harmonic resonance, and is absorbed at higher $w_\text{eff}^2$.

It can also be seen in Fig. \ref{fig:discussion_top_side_weff} that $N_{\parallel}$ is larger along the top launch ray trajectory. Eq. \ref{eq:relativistic} shows
that the width of the allowed power deposition region, as determined by relativistic effects, is larger when $N_\parallel^2$ is larger. The value of $N_\parallel^2$ at the launching point is determined mainly by the toroidal launch angle, and is larger when the toroidal launch angle is larger.

\section{\label{sec:conclusions} Discussion}

Calculations in recent years have suggested that nonlinear effects can significantly impact the stabilization of magnetic islands by RF driven currents. \cite{reiman_suppression_2018,rodriguez_rf_2019,rodriguez_rf_2020,jin_pulsed_2020,frank_generation_2020,reiman_disruption_2021,jin_two-fluid_2021,nies_calculating_2020,jin_hot_ion_2021} The effects can potentially be used to improve stabilization efficiency. If not properly accounted for in the aiming of ray trajectories, they may lead to a shadowing effect that impairs stabilization. Calculations also suggest that a broad RF driven current in a fusion reactor could provide automatic stabilization of tearing modes, with the RF driven current coalescing in a magnetic island when the island reaches sufficient width.

Quantitative validation of the theory predicting RF condensation has, however, been lacking. This paper has investigated some possibilities for testing and refining the theory on a contemporary experimental device. The paper has focused primarily, but not exclusively, on electron cyclotron (EC) waves launched from above the plasma (top launch). Top launch provides increased electron cyclotron current drive (ECCD) efficiency relative to outside launch, \cite{chen2021doubling,chen_doubling_2022}  and is therefore more likely to be used for a fusion reactor, where recirculating power can have a major impact on the cost of electricity. Top launch ECCD is also most efficient at large toroidal launch angles, where nonlinear effects are most pronounced. Outside launch ECCD (launched from the low field side of the plasma), in contrast, is most efficient at intermediate toroidal launch angles. In the ASDEX Upgrade (AUG) calculations described here, the current drive efficiency peaks at a toroidal launch angles of about $20\degree$ relative to the $-\hat{R}$ direction. The largest possible toroidal launch angle for the AUG launchers relative to the $-\hat{R}$ direction is about $25\degree$. In contrast, the toroidal launch angle relative to $-\hat{R}$ in the DIII-D top launch experiments was $61\degree$ .

 ECCD island stabilization experiments have thus far largely been done using outside launch, with toroidal launch angles that maximize the current drive efficiency. The ASDEX upgrade (AUG) calculations described in this paper find that the nonlinear effects are predicted to be weak for such toroidal launch angles. It is not surprising, then, that the nonlinear effects discussed in this paper have not made themselves known in the analysis of such experiments, although Bardoczi and Logan have reported seeing a signature of RF condensation\cite{bardoczi21}. In a reactor with top launch ECCD, the nonlinear effects are predicted to be more significant.

Our work shows that, in the DIII-D calculations, the combined effects of stiffness above the microinstability threshold and rotation can lead to a muted and more gradual onset of the nonlinear effects. With an increased fraction of the power deposited in the island periphery, the local microinstability threshold is encountered first in the peripheral region, with the boundary of the stiff region moving gradually inward. The temperature increase in the island is more gradual. There is no indication that a bifurcation threshold is being approached.
 
 These calculations may overstate the effect of stiffness. With a flat density profile, it would be expected that the ITG threshold will be encountered when the ion temperature gradient becomes sufficiently large. There is evidence that this would affect the ion temperature gradient directly, but would only affect the electron temperature gradient through the coupling of the electrons to the ions. It has been found on both W7X and AUG that, for a plasma heated by ECH, the ion temperature is clamped when it reaches the ITG threshold, but the electron temperature can significantly exceed the ion temperature, depending on the coupling between the electrons and the ions. \cite{beurskens21,beurskens22} Calculations of these two-fluid effects are beyond the scope of this paper.
 
 The picture of the combined impact of rotation and stiffness suggests that, for rotating islands, it may be advantageous to use modulated ECCD to deposit the power more effectively in the central portion of the island and retain some of the advantages that are seen with locked islands. In particular, in a validation experiment, locking may pose a problem for ECE measurements, depending on the location of the line of sight of the ECE diagnostic relative to the EC launchers. It may be preferable to prevent the island from locking by feed forward entrainment in a rotating resonant magnetic perturbation.\cite{choi_feedforward_2018} For a rotating island, it will be of interest to calculate the predicted impact on the nonlinear effects of modulation with various duty cycles. These calculations are beyond the scope of this paper.

 For a fusion reactor with a broad RF driven current, the conventional linear theory of the stabilization of islands via unmodulated ECCD predicts little or no stabilization of islands unless the aiming of the RF is appropriately adjusted by external feedback control. The linear stabilization effect relies on a current density gradient across the island that decreases from the island center radially outward. There is a geometric effect such that equilibration of this current along magnetic field lines yields a stabilizing resonant component of the current density. The nonlinear RF condensation effect would, however, automatically lead to coalescence of the ECCD in the island, producing a stabilizing resonant component of the field without the need for feedback control. As described above, depending on the strength of the coupling between the ions and electrons (i.e. depending on the density and temperature), the combined effects of stiffness and rotation may lead to a muted and gradual onset of the nonlinear effects. The nonlinear effect may nevertheless be sufficient to stabilize the island. Prediction of the magnitude of the stabilizing resonant current in this case will require the application of a quantitative model that has been benchmarked against experimental data. The purpose of this paper is to investigate possible scenarios for providing that benchmarking. 
 
 The stabilization could be improved by modulating the ECCD with a frequency and phase determined by the Mirnov oscillations as soon as such oscillations are detected. That would require an increased level of EC power in order to produce the same current density as the unmodulated EC. A broad ECCD profile would then provide some linear stabilization, and the stabilization could be enhanced by a condensation effect that is comparable to that for a locked island.

 A growing island that is not stabilized in a fusion reactor will likely lock to the wall at relatively small island width. Future large tokamaks are expected to rotate much more slowly than DIII-D or AUG, and the islands are expected to lock at correspondingly smaller widths. It has been projected that the $q=2$ island in the ITER base case will lock at a width of less than $5\%$ of the minor radius.\cite{lahaye17} ECCD stabilization of locked islands has been demonstrated in a series of experiments on DIII-D by Volpe \textit{et al}. \cite{volpe_2015}.
 To stabilize locked islands, it will be necessary for the islands to lock at a phase where the EC launcher is approximately aligned with the island O-point. \cite{nies_calculating_2020} This can be arranged by a proper adjustment of the field error compensation coils. As in contemporary tokamaks, ITER will have field error compensation coils that substantially reduce the magnitude of undesirable nonaxisymmetric resonant components of the magnetic field produced by finite tolerances in the placement of the magnetic field coils. It is expected that future large tokamaks will also have such coils. The phase of locked islands will be determined by the residual resonant components of the field after partial cancellation. A slight adjustment of the field produced by the field error correction coils, produced by a slight adjustment in the relative magnitudes of the currents in the coils, will be sufficient to lock the islands at the desired phase. If the island is locked at the appropriate phase, a broad ECCD profile would then produce a linear stabilizing effect, which would be enhanced by RF condensation.

This paper has focused on possible scenarios for testing and refining the theory predicting a nonlinearly enhanced temperature perturbation in an island. The theory describing the electron cyclotron power deposition and  current drive in a plasma with a given temperature and density has been well validated. \cite{prater_heating_2004}. The temperature perturbation in the island produced by a given level of power deposition, and the resulting nonlinear enhancement of the power deposition,  depends on heat transport in the island, for which there is at present only a limited quantitative understanding. There is strong evidence that the cross-field heat transport in the island is small relative to the turbulent heat tranport outside the island when the temperature gradient in the island is sufficiently small. \cite{inagaki_diffusivity_2004,spakman2008,bardoczi_diffusivity_2016,hornsby2011} It is presumed that the heat transport in the island rises to the same level as that outside the island when the temperature gradient in the island gets sufficiently large. There is at present only a limited quantitative understanding of this transition. A set of experiments along the lines described in this paper would be a step towards providing that information.

Although we have focused on the issue of the nonlinearly enhanced temperature perturbation in this paper, we are ultimately interested in that effect for the purpose of stabilizing islands via RF driven currents. That stabilizing effect is stronger than might be surmised by considering only the temperature perturbation. If the temperature perturbation increases with increasing EC power until it reaches a critical temperature gradient above which the temperature profile in the island becomes stiff, the EC driven current will nonetheless continue to increase with increasing EC power. Any nonlinear narrowing of the power deposition profile that occurs below the temperature gradient threshold will continue to be reflected in the current density profile as the driven current increases further. For a broad EC current profile, the helical perturbation of the temperature profile will continue to be reflected in a resonant perturbation of the current density, with the amplitude of that resonant current density perturbation continuing to increase in proportion to any further increase in the EC power even if the temperature perturbation is clamped by profile stiffness.



\section{Acknowledgments}
Work supported by DOE contracts DE-SC0023236, DE-AC02-09CH11466, and DE-FG02-97ER54415. This material is based upon work supported by the U.S. Department of Energy, Office of Science, Office of Fusion Energy Sciences, using the DIII-D National Fusion Facility, a DOE Office of Science user facility, under Award(s) DE-FC02-04ER54698, DE-AC52-07NA27344, and DE-SC0022270.

This work has been carried out within the framework of the EUROfusion Consortium, funded by the European Union via the Euratom Research and Training Programme (Grant Agreement No 101052200 — EUROfusion). Views and opinions expressed are however those of the author(s) only and do not necessarily reflect those of the European Union or the European Commission. Neither the European Union nor the European Commission can be held responsible for them.

\section{Disclaimer}
This report was prepared as an account of work sponsored by an agency of the United States Government. Neither the United States Government nor any agency thereof, nor any of their employees, makes any warranty, express or implied, or assumes any legal liability or responsibility for the accuracy, completeness, or usefulness of any information, apparatus, product, or process disclosed, or represents that its use would not infringe privately owned rights. Reference herein to any specific commercial product, process, or service by trade name, trademark, manufacturer, or otherwise does not necessarily constitute or imply its endorsement, recommendation, or favoring by the United States Government or any agency thereof. The views and opinions of authors expressed herein do not necessarily state or reflect those of the United States Government or any agency thereof.
\begin{appendices}
\section{Appendix A: Approximate $w_\text{eff}^2$ formula and its accuracy}\label{app:weff_goodness}

The $w_\text{eff}^2$ values presented in this paper are approximate values\cite{nies_calculating_2020} given by:
\begin{equation}
-w_\text{eff}^2 = \mu\left(1-\frac{2\Omega/\omega}{1-N_{\parallel}^2}\right)+\frac{\xi_2I_{3/2}(\xi_2)}{I_{5/2}(\xi_2)}-\frac{5}{2},
\end{equation}
where

\begin{equation}
R_2 = \sqrt{\left(\frac{2\Omega}{\omega}\right)^2-1+N_{\parallel}^2},
\end{equation}

\begin{equation}
\xi_2 = \frac{N_{\parallel}R_2\mu}{1-N_{\parallel}^2}.
\end{equation}

Using the cases with scaled $B_\text{tor}$ in \ref{sec:discussions}, we show that the formula closely approximates the numerical values evaluated using $w_\text{eff}^2 \equiv T\partial_T[ln (dP_{lin}/ds)]$ using finite differences.

\begin{figure}[h]
     \centering
     \begin{subfigure}[b]{0.49\textwidth}
         \centering
         \includegraphics[width=\textwidth]{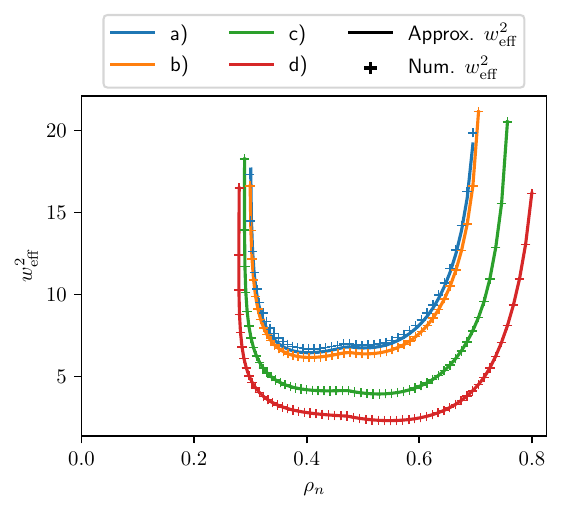}
         \caption{Numerical and approximate $w_\text{eff}^2$ along trajectories for case (a) through (d) in Fig. \ref{fig:discussion_stages}.}
         \label{fig:appendix_weff_traj}
     \end{subfigure}
     \hfill
     \begin{subfigure}[b]{0.49\textwidth}
         \centering
         \includegraphics[width=\textwidth]{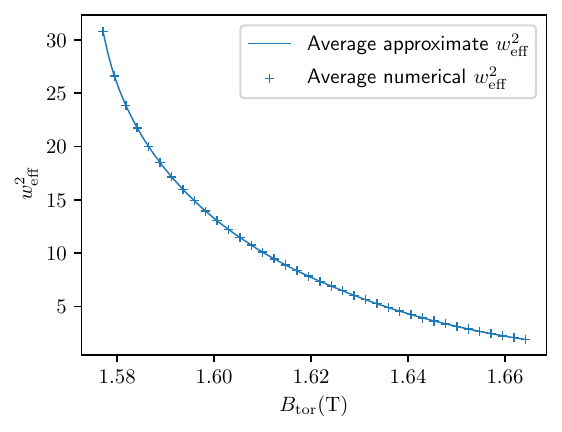}
         \caption{Trajectory-averaged numerical and approximate $w_\text{eff}^2$ across the $B_\text{tor}$ sweep in Fig. \ref{fig:discussion_stages}.}
         \label{fig:appendix_weff_avg}
     \end{subfigure}
        \caption{Comparison between numerical and approximate $w_\text{eff}^2$.}
        \label{fig:appendix_weff}
\end{figure}

\end{appendices}

\bibliographystyle{unsrt}  
\bibliography{article.bib}  

\end{document}